\documentclass[showpacs,aps,psfig,epsf]{revtex4}
\usepackage{amssymb}

\usepackage{graphicx}

\begin{document}

\title{Charge imbalance and Josephson effects in superconductor-normal metal
mesoscopic structures. }
\author{A.F. Volkov$^{1,2}$.}

\address{$^{(1)}$Theoretische Physik III,\\
Ruhr-Universit\"{a}t Bochum, D-44780 Bochum, Germany\\
$^{(2)}$Institute for Radioengineering and Electronics of Russian Academy of\\
Sciences,11-7 Mokhovaya str., Moscow 125009, Russia\\}

\begin{abstract}
We consider \ a $SBS$ Josephson junction the superconducting electrodes $S$
of which are in contact with normal metal reservoirs ($B$ means a barrier).
For temperatures near $T_{c}$ we calculate an effective critical current $%
I_{c}^{\ast }$ and the resistance of the system at the currents $I<$ $%
I_{c}^{\ast }$ and $I>>I_{c}^{\ast }$. It is found that the charge
imbalance, which arises due to injection of quasiparticles from
the $N$ reservoirs into the $S$ wire, affects essentially the
characteristics of the structure. The effective critical current
$I_{c}^{\ast }$ is always larger than the critical current $I_{c}$
in the absence of the normal reservoirs and increases with
decreasing the ratio of the length of the $S$ wire $2L$ to the
charge imbalance relaxation length $l_{Q}$. It is shown that a
series of peaks arises on the $I-V$ characteristics due to
excitation of the Carlson-Goldman collective modes. We find the
position of Shapiro steps which deviates from that given by the
Josephson relation.
\end{abstract}

 \maketitle

\bigskip

\section{Introduction}

In the theory of the Josephson effect in weak links \cite{Josephson} it is
assumed that the superconducting electrodes are in equilibrium. This means,
in particular, that a gauge-invariant potential $\mu $

\begin{equation}
\mu =\frac{1}{2}\partial \chi /\partial t+eV,  \label{Mu}
\end{equation}
is equal to zero \cite{Ft1}(we set $\hbar =1$). The Josephson
relation follows immediately from this condition

\bigskip
\begin{equation}
2e(V_{2}-V_{1})=\partial \varphi /\partial t,  \label{JosRelation}
\end{equation}%
where $(V_{2}-V_{1})$ is the voltage drop across the Josephson junction and $%
\varphi =\chi _{1}-\chi _{2}$ is the phase difference. Most works
on studies of the Josephson effects were carried out under
equilibrium conditions so that the potential $\mu $ is zero in the
superconducting electrodes $S_{1,2}$, and Eq. (\ref{JosRelation})
is satisfied \cite{Kulik,Barone}. The potential $\mu $ is related
to a so called charge imbalance arising due to a different
population of the electronlike and holelike branches of the
quasiparticle spectrum of a superconductor \cite{Ft2}. The
conditions under which this
charge imbalance may arise were studied experimentally \cite%
{Cl,Yu,KlMoi,Dolan,VanHar,Mamin,Chien,Strunk} and theoretically \cite%
{Rieger,Tin,SSJLTP,AV,AVUsp,Ovch}.

In the last decade a great progress in preparation of
superconductor/normal metal ($S/N$) nanostructures has been
achieved. Varies properties of these
structures were studied experimentally. One can mention the study of transport %
\cite{Petrashov,Pothier,Charlat,Chien,Klapwijk,Shapira} and thermoelectric %
\cite{ChandraTherm,SosninTherm} properties, the measurements of
the density-of-states \cite{EsteveDOS} etc. Recently a transition
of a thin and short superconducting wire into a resistive state
caused by a current $I$ was investigated \cite{Bezryadin}. In
particular, an increase of the critical current $I_{c}$ has been
observed when an external magnetic field $H$ is applied
\cite{Bezryadin}. A possible reason for the observed increase of
$I_{c}$ may be a polarization of magnetic impurities by the
magnetic field $H$ \cite{Feigelman,Bezryadin}. Another mechanism
has been suggested in Ref.\cite{Vodolazov}. It has been proposed
that an increase of the critical currents characterizing
phase-slip centers in a thin $S$ wire may be due to a shortening
of the charge imbalance relaxation length $l_{Q}$ under the
influence of $H$. However, it remains unclear how this mechanism
can work in a pure superconducting state when there is no charge
imbalance. As is known (for review see
\cite{AVUsp,Rammer,Kopnin}), the charge imbalance arises only
under nonequilibrium conditions. For example, this mechanism may
be realized in a superconductor-normal metal ($SN$) nanostructure
with a bias current, where the charge imbalance is built up due to
injection of
quasiparticles into the $S$ wire from the $N$ reservoirs \cite%
{Rieger,Tin,SSJLTP,AV,AVUsp,Ovch}.

In the present, paper we investigate the Josephson effect in a
$NSBSN$
structure under nonequilibrium conditions, in the presence of a bias current $%
I$ flowing through $SN$ interfaces.\ In this case the potential $\mu $
arises due to injection of quasiparticles into the superconductors $S$ from
the normal reservoirs $N$ ($B$ is a barrier of an arbitrary type). We show
that if the length of the $S$ wire $2L$ is less than or comparable with the
charge imbalance relaxation length $l_{Q},$ the critical current increases
and may significantly exceed the critical current $I_{c}$ in the absence of
the normal reservoirs. The resistance of the system also will be calculated
for the currents $I\leq I_{c}^{\ast }$ and $I\gg I_{c}^{\ast }$, where $%
I_{c}^{\ast }$ is an effective critical current. We find the position of
Shapiro steps and show that peaks associated with the excitation of
collective modes of the Carlson-Goldman type arise on the $I-V$
characteristics.

Note that in recent publications \cite{Keizer,Peeters} a $NSN$
system without the Josephson junction was considered and the $I-V$
curve of the system was calculated numerically.

\section{Model and Basic Equations}

We consider the structure shown in Fig.1. We assum that the
barrier is located in the middle of the $S$ wire, but the results
remains qualitatively unchanged for a system with the barrier
located at the $S/N$ interface. The system consists of a
superconducting wire ($S$ wire) connecting two normal reservoirs
$N$. In the middle of the $S$ wire there is a barrier which
provides the Josephson coupling. The transparency of the $%
SN$ interface is assumed to be high. The analysis can be easily
generalized to the case of arbitrary $NS$ interface
transparencies. We also assume  that the temperature $T$ is close
to the critical one $T_{c}$ and therefore the inequality

\begin{equation}
\Delta \ll T  \label{Tdelta}
\end{equation}%
is satisfied ($\Delta $ is the order parameter in the $S$ wire). In this
case the effects of the branch imbalance are most significant. The lengths $%
L $ and $l_{Q}$ are assumed large in comparison with the
Ginsburg-Landau correlation length $\xi _{GL}$, i.e.,

\begin{equation}
\xi _{GL}\approx 1.2\sqrt{D/T_{c}}(T_{c}/\Delta )\ll \{L,l_{Q}\},
\label{KsiGL}
\end{equation}%
where $D$ is the diffusion coefficient in the $S$ wire. The charge
imbalance relaxation length $l_{Q}$ is determined by inelastic
scattering processes and may be rather long
(\cite{Tin,SSJLTP,AV,Ovch,AVUsp}). This assumption
allows us to consider the order parameter constant in the major part of the $%
S$ wire. The same limit was considered in Refs. \cite{AV,Ovch} where the
resistance of the superconductor in a $SN$ system was calculated. The
relation between $L$ and $l_{Q}$ may be arbitrary.

\begin{figure}
\begin{center}
\includegraphics[width=0.9\textwidth]{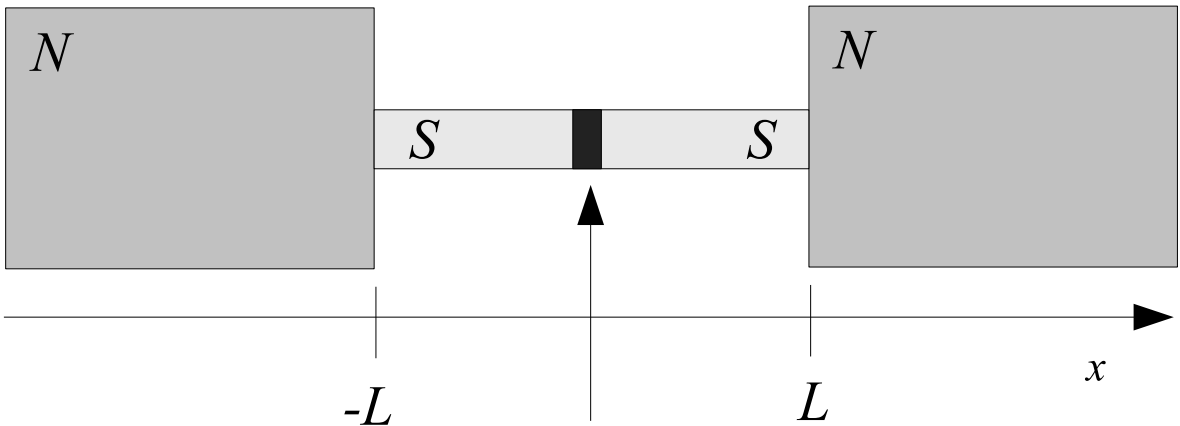}
\end{center}
\large{\caption{Schematic view of the system under
consideration.}}
\end{figure}
Our aim is to obtain a relationship between the current $I$ and the voltage $%
V_{L}\equiv V(x)$ at $x=L$ (the voltage difference between the $N$
reservoirs is $V_{NN}=-2V_{L}$). We restrict ourselves with
voltages $V_{NN}$ small compared to the energy gap $\Delta :$
$eV_{NN}\ll \Delta $. In this
limit the distribution function (longitudinal in terms of Ref. \cite{SSJLTP}%
), which determines the order parameter $\Delta ,$ is close to the
equilibrium one. Another distribution function denoted by $f_{1}$ in Ref. %
\cite{LO} (transverse in terms of Ref. \cite{SSJLTP}) was found in Ref. \cite%
{SSJLTP,AV,Ovch}. To be more exact, the so called anomalous Green's
function, $\hat{g}^{(a)}=$ $\hat{g}^{R}\hat{\tau}_{3}f_{1}-f_{1}\hat{\tau}%
_{3}\hat{g}^{A}$, was found in Refs. \cite{AV,Ovch,AVUsp}, where $\hat{g}%
^{R(A)}$ are the retarded (advanced) Green's functions. Using the functions $%
\hat{g}^{(a)}$ and $\hat{g}^{R(A)}$, one can obtain the expression
for the current $I$ in the $S$ wire, which in the main
approximation in the parameters $\Delta /T$ and $V/T$ is equal to
\begin{equation}
I=\mathcal{S}\sigma (E+\frac{\pi \Delta ^{2}}{2Te}Q),  \label{Icurrent}
\end{equation}%
where $\mathcal{S}$ is the cross section area of the $S$ wire, $\sigma $ is
the conductivity of the $S$ wire in the normal state. We assume that the
cross section area is small compared to $\{\xi _{GL}^{2},\lambda _{L}^{2}\}$
so that all the vectors $\{\mathbf{I,E,Q}\}$ have only the $x-$component $%
\{I,E,Q\}$ and depend on $x$. The electric field $E=-\partial V(x)/\partial
x $ (we drop the vector potential $A$ because we consider only the
longitudinal electric field choosing a corresponding gauge). The condensate
momentum $Q$ is defined as

\begin{equation}
Q=(1/2)\partial \chi /\partial x-2\pi A/\Phi _{0}  \label{Q}
\end{equation}
where $\Phi _{0}=hc/2e$ is the magnetic flux quantum; the vector potential $%
A $ can be dropped because we do not consider the action of magnetic field.
The momentum $Q$ obeys the equation

\begin{equation}
\partial Q/\partial t=eE+\partial \mu /\partial x,  \label{Qeq}
\end{equation}

\bigskip The spatial and temporal variation of the gauge-invariant potential
$\mu (x,t)$ is described by the equation \cite{AVUsp,SchoenRev}

\begin{equation}
(\partial /\partial t+\gamma )\mu =v_{CG}^{2}\partial Q/\partial x,
\label{MuEq}
\end{equation}%
where $\gamma $ is a quantity which determines the charge
imbalance relaxation rate. It is related to inelastic scattering
processes ($\gamma =1/\tau _{\epsilon }$, where $\tau _{\epsilon
}$ is the inelastic relaxation time), the condensate momentum
($\gamma \sim DQ^{2}\Delta /T$) in the
presence of condensate flow or the gap anisotropy \cite%
{Tin,SSJLTP,AV,Ovch}. The velocity $v_{CG}=\sqrt{2D\Delta }$ is the velocity
of propagation of the Carlson-Goldman collective mode \cite%
{CG,SSCollModes,AVCollModes,AVUsp,SchoenRev}. In the stationary case Eq.(\ref%
{MuEq}) can be written in the form

\begin{equation}
l_{Q}^{-2}\mu =(e/\sigma )\partial j_{S}/\partial x=\partial ^{2}\mu
/\partial x^{2},  \label{EqMuSt}
\end{equation}
where $j_{S}=\sigma (\pi \Delta ^{2}/2Te)Q$ is the density of the
supercurrent; $l_{Q}^{{}}=\sqrt{4TD/(\pi \gamma \Delta )}$ is the
penetration depth of the electric field (or the charge relaxation length).
Eq.(\ref{EqMuSt}) describes the conversion of the quasiparticle and
superconducting currents. One can see that the potential $\mu $ arises if
the divergence of the supercurrent (or quasiparticle current) differs from
zero.

The current in the $S$ wire can also be written as the current through the
Josephson junction. In the main approximation it is equal to (see Appendix)

\begin{equation}
I=-\frac{2V_{0}}{R_{B}}+I_{c}\sin \varphi ,  \label{CurrentJJ}
\end{equation}%
where $2V_{0}\equiv 2V(0+)=-2V(0-)$ is the voltage drop across the
Josephson junction ($V_{0}$ is negative), $R_{B}$ is the barrier
resistance, $I_{c}$ is the Josephson critical current, and
$\varphi =\chi (0_{+})-\chi (0_{-})$ is
the phase difference. For simplicity we drop here the displacement current $%
C\partial V_{0}/\partial t$ assuming that the parameter $%
(CR_{B})e(I_{c}R_{B})$ is small (the main results concerning the critical
current and the resistances of the system do not depend on the presence of
the displacement current). At $V(L)\ll \Delta $ the critical current equals $%
I_{c}=\pi \Delta ^{2}/4TeR_{B}.$ The first term in
Eq.(\ref{CurrentJJ}) is the quasiparticle current and the second
term is the supercurrent.

Eqs.(\ref{Icurrent}-\ref{CurrentJJ}) describe the system. These
equations should be complemented by boundary conditions. Since we
consider the voltages smaller than $\Delta $, the distribution
function $f_{l},$ which determines the supercurrent, is close to
the equilibrium one: $f_{l}=$ $\tanh (\epsilon /2T)$. This implies
the conservation of the quasiparticle (correspondingly
superconducting) currents at the Josephson junction, i.e.,

\begin{equation}
\mathcal{S}\sigma E(0_{+})=-\frac{2V_{0}}{R_{B}}=I-I_{c}\sin \varphi .
\label{BC1}
\end{equation}

Another boundary condition relates the electric field at the edge of the $S$ wire $E(L)$ to the electric field
in the $N$ region $E_{N}$. As is shown in Refs.\cite{AV,Ovch}, the
electric field $E(x)$ experiences a jump at the $SN$ interface

\begin{equation}
\lbrack E]_{SN}\equiv E_{N}-E(L)=rE(L)  \label{BCforE}
\end{equation}%
where $r\approx 0.7((\Delta /T)l_{\epsilon }/\xi
_{GL})^{1/4}\approx 1.17[((T_{c}-T)/T_{c})^{2}\tau _{\epsilon
}/\tau ]^{1/4}$; $\tau _{\epsilon },\tau $ are the inelastic and
elastic scattering times. This jump is small if the temperature
$T$ is close to $T_{c}$ and therefore the electric field
(accordingly the quasiparticle current) is continuous at the $SN$
interface. For a finite value of $r$ and equal conductivities in
the $S$ and $N$ regions we get from Eq.(\ref{BCforE})

\begin{equation}
E(L)=E_{N}(1-\frac{r}{1+r})  \label{BC2}
\end{equation}%
Here the second term on the right is small near $T_{c}$. Solving Eqs.(\ref%
{Icurrent},\ref{Qeq},\ref{MuEq},\ref{CurrentJJ}) with the boundary
conditions (\ref{BC1}-\ref{BC2}), we can calculate, in particular,
the
effective critical current $I_{c}^{\ast }$ and the resistance of the system $%
R$.

\section{Resistance and Josephson critical current.}

\bigskip Eqs.(\ref{Icurrent},\ref{Qeq},\ref{MuEq},\ref{CurrentJJ}) are
nonlinear equations in partial differentials in which all
functions depend on time $t$ and coordinate $x$. Therefore the
solution to these equations can not be obtained in a general case.
We consider limiting cases of small and large currents $I$.

\textit{a) Stationary case; }$I<I_{c}^{\ast }.$

If the current does not exceed a critical value $I_{c}^{\ast }$
(the effective critical current $I_{c}^{\ast }$ will be found
below), the phase difference $\varphi $ and other quantities do
not depend on time. In particular, $\partial Q/\partial t=0$ and
the electric field is

\begin{equation}
eE(x)=-\partial \mu /\partial x  \label{StatE}
\end{equation}%
The momentum of the condensate is found from Eq.(\ref{Icurrent})

\begin{equation}
\frac{\pi \Delta ^{2}}{2T}Q=eIR_{n}+\partial \mu /\partial x
\label{StatQ}
\end{equation}%
where $R_{n}=(\mathcal{S}\sigma )^{-1}$ is the resistance of the $S$ wire in
the normal state per unit length. The potential $\mu $ is described by Eq.(%
\ref{EqMuSt}). At small currents when the condensate flow does not
contribute to the relaxation of the charge imbalance, the length
$l_{Q}^{{}}$ equals: $l_{Q}=\sqrt{4(D\tau _{\epsilon })T/\pi
\Delta }$ \cite{SSJLTP,AV,Ovch}. The solution of Eq.(\ref{EqMuSt})
is

\begin{equation}
\mu (x)=A\cosh (x/l_{Q})+B\sinh (x/l_{Q})  \label{Mux}
\end{equation}%
The electric field and the potential $V(x)$ equal
\begin{equation}
eE(x)=-[A\sinh (x/l_{Q})+B\cosh (x/l_{Q})]/l_{Q},\text{ \ }eV(x)=A\cosh
(x/l_{Q})+B\sinh (x/l_{Q})+V_{\ast }  \label{E+V}
\end{equation}%
where $V_{\ast }$ is an integration constant.

First we neglect the second term on the right in Eq.(\ref{BC2}).
From Eqs.(\ref{BC1},\ref{BC2}) we find the coefficient $B$ and $A$

\begin{equation}
B=-eR_{Q}(I-I_{c}\sin \varphi ),\text{ \ }A=-(eR_{Q}I+BC)/S  \label{BA}
\end{equation}
where $R_{Q}=l_{Q}/(\mathcal{S}\sigma )$, $C\equiv \cosh \theta$ and $%
S\equiv \sinh \theta$ with $\theta = L/l_{Q}$.

From Eq.(\ref{E+V}) we obtain

\begin{equation}
eV_{L}=eV_{0}+A(C-1)+BS
\end{equation}%
Writing $V_{0}$ in terms of $B$ (see Eqs.(\ref{BC1},\ref{BA}) ), we can
represent $V_{L}$ in the form

\begin{equation}
eV_{L}=B(\frac{b}{2}+\frac{C-1}{S})-IR_{Q}\frac{C-1}{S}  \label{V_L}
\end{equation}

\bigskip In the stationary case one has $eV_{0}=\mu (0)=A.$ Combining Eqs.(%
\ref{BC1},\ref{BA}), we obtain from this formula

\begin{equation}
I(\frac{b}{2}+\frac{C-1}{S})=I_{c}(\frac{b}{2}+\frac{C}{S})\sin \varphi
\label{EffI_c1}
\end{equation}
where $b\equiv R_{B}/R_{Q}$. Eq.(\ref{EffI_c1}) determines the effective
critical current $I_{c}^{\ast }$

\begin{equation}
I_{c}^{\ast }/I_{c}=\frac{bS+2C}{bS+2(C-1)}=\frac{b\tanh (\theta
/2)+1+\tanh ^{2}(\theta /2)}{\tanh (\theta /2)(b+2\tanh (\theta
/2))}  \label{EffI_c}
\end{equation}

It is seen that the effective critical current $I_{c}^{\ast }$ is
always larger than the critical current $I_{c}$ in the absence of
the charge imbalance. In the limit of large $b=R_{B}/R_{Q}$ or
$\theta =L/l_{Q}$ the effective critical current coincides with
$I_{c}$. Therefore in a Josephson junction with a weak coupling
between superconductors ($b\gg 1$), the effects of charge
imbalance are not important. For a short $S$ wire, we obtain

\begin{equation}
I_{c}^{\ast }/I_{c}=\frac{b\theta +2}{\theta (b+2\theta )}  \label{EffI_c3}
\end{equation}%
One can see that $I_{c}^{\ast }$ diverges at $\theta \rightarrow
0$. This fact is in agreement with the results of
Ref.\cite{Zaitsev}, where a very short $NSS^{\prime }$ system was
considered, and it was concluded that the stationary state exists
at any currents $I.$ However our consideration is valid only for
not too small $\theta $ ($\theta =L/l_{Q}\succsim \xi
_{GL}/l_{Q}$). The dependence of $I_{c}^{\ast }/I_{c}$ on $\theta
$ is shown in Fig.2 for different $b.$
\begin{figure}
\begin{center}
\includegraphics[width=0.9\textwidth]{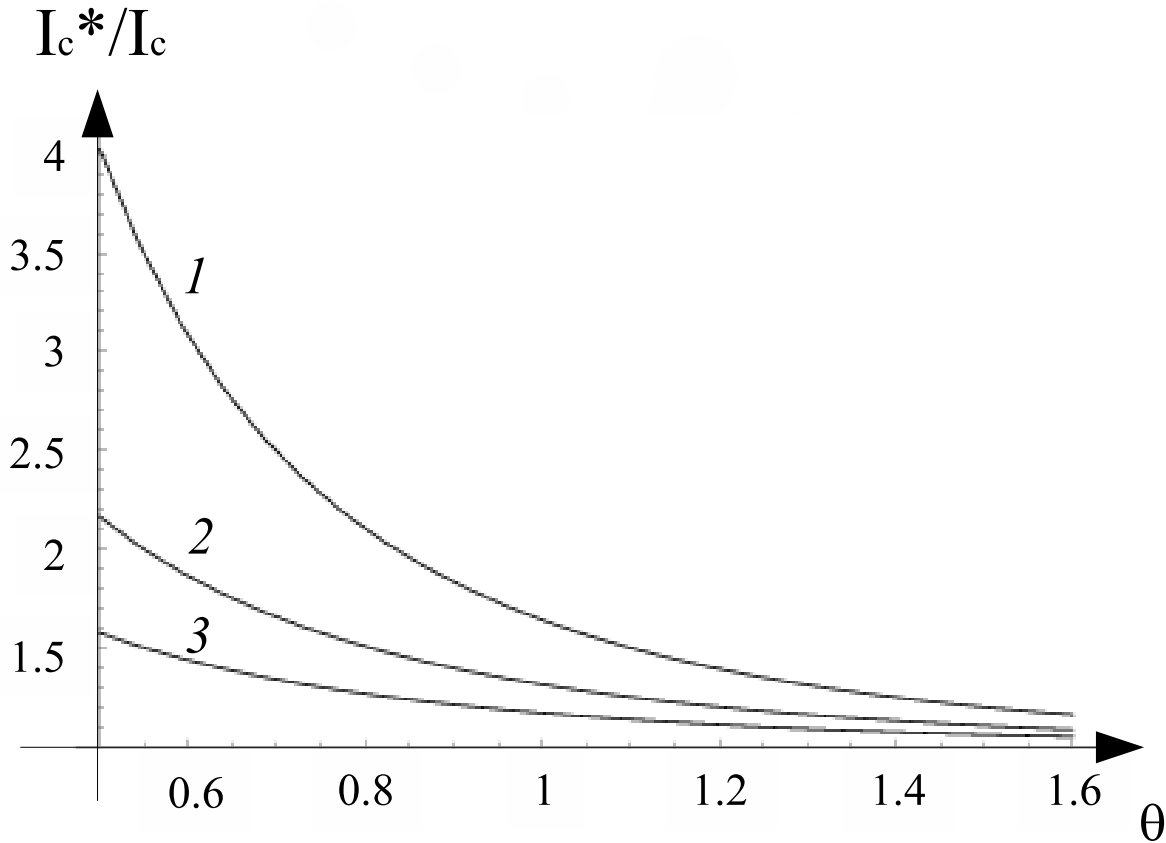}
\end{center}
\large{\caption{The normalized effective critical current as a
function of $\theta=L/l_{Q}$ for $b=R_{B}/R_{Q}$ = 0.2 (curve 1),
$b$ = 2 (curve 2), and $b$ = 5 (curve 3) }}
\end{figure}

What is the reason for the increase of the effective critical
current from the physical point of view? The critical current is
defined as a maximum current $I_{\max }$\ at which the stationary
state is possible: $\partial \varphi /\partial t=0.$\ The current
$I_{\max }$\ in an ordinary equilibrium Josephson junction
coincides with the critical current $I_{c}$\ because the
first term in Eq.(\ref{CurrentJJ}) in the stationary state is zero: -$%
2V_{0}=(\hbar /2e)\partial \varphi /\partial t=0.$\ As we noted in
the Introduction, this equation follows from the fact that the
potential $\mu (x)$\ is zero. The same situation takes place near
the $S/B/S$ interfaces in a long ($L>>l_{Q}$) $S$ wire. The charge
imbalance, which determines $\mu (x)$\, arises only in the
vicinity of $S/N$ interface and decays on the scale of order
$l_{Q}$. Therefore the effective critical current is the maximum
current at which the stationary state with nonzero potentials $V$\
and $\mu $\ is possible. Here we find this current for the case of
a short system: $L<<l_{Q}$. Then, the electric field $E(x)$\ and
potential $V(x)$\ can be written as

\begin{equation}
eE(x)\cong -[A(x/l_{Q})+B]/l_{Q},eV(x)\cong A+B(x/l_{Q})\text{ }
\label{EshortL}
\end{equation}%

The first term in the square brackets of Eq.(\ref{EshortL})
describes the conversion of the quasiparticle current into the
condensate current. In the normal state it is zero because
$l_{Q}\rightarrow \infty $. \ At $x=L$\, the charge is transferred
by quasiparticles (we ignore the jump in the electric field
setting $r=0$): $S\sigma E(L)=I$. At $x=0$\ the quasiparticle
current is $S\sigma E(0_{+})=I-I_{c}\sin \varphi .$\ Therefore the
coefficient $A$\ related to the conversion of the quasiparticle
current into the condensate one is equal to\

\begin{equation}
-A(L/l_{Q})\cong eI_{c}R_{Q}\sin \varphi .\text{ } \label{AshortL}
\end{equation}%
On the other hand, at currents $I$\ large in comparison with
$I_{c}$\, one has

\begin{equation}
I\cong -2V_{0}/R_{B}\cong -2A/eR_{B}.  \label{IshortL}
\end{equation}

Thus, from Eqs. (\ref{AshortL}-\ref{IshortL}), we find the maximum
current for the stationary state

\begin{equation}
I_{c}^{\ast }\cong I_{c}(2l_{Q}/bL)
\end{equation}

This formula is valid if the ratio $(l_{Q}/bL)$\ is large. If the
current exceeds this value, the stationary state is not possible.
One can say that the length of the S wire is too short to provide
the conversion of the quasiparticle current into the condensate
current $I_{c}\sin \varphi $.

If we take into account the second term in the boundary condition (\ref{BC2}%
), we obtain for $I_{c}^{\ast }$

\begin{equation}
I_{c}^{\ast }/I_{c}=\frac{2+b\tanh \theta }{\tanh \theta \lbrack b+2\tanh
(\theta /2)+2r/(S(1+r))]}  \label{I_Cr}
\end{equation}

The resistance of the system $R_{0}=V_{NN}/I=|2V_{L}|/I$ at $I<I_{c}^{\ast }$
can be easily found with the aid of Eqs. (\ref{V_L},\ref{EffI_c1})

\begin{equation}
R_{0}=2R_{Q}\frac{2\tanh \theta +b}{2+b\tanh \theta }  \label{R_0}
\end{equation}
In limiting cases we find

\begin{equation}
\text{ }R_{0}=2R_{Q}\left\{
\begin{array}{ll}
(2\theta +b)/(2+b\theta ), & \theta \ll 1 \\
1, & \theta \gg 1%
\end{array}
\right.
\end{equation}

\bigskip The resistance $R_{0}$ is equal to $R_{B}$ at $\theta \rightarrow 0$
and to $2R_{Q}$ at $\theta \rightarrow \infty $. Thus, the
resistance of a
short system is a combination of the barrier resistance $R_{B}$ and $%
R_{Q} $ (the resistance of the $S$ wire near the $SN$ interface). The
resistance of a long system equals the resistance of the $S$ wire near the $%
SN$ interface on the scale $l_{Q}.$ The dependence of the
resistance $R_{0} $ on $\theta $ for different $b$ is shown in
Fig.3.

\begin{figure}
\begin{center}
\includegraphics[width=0.9\textwidth]{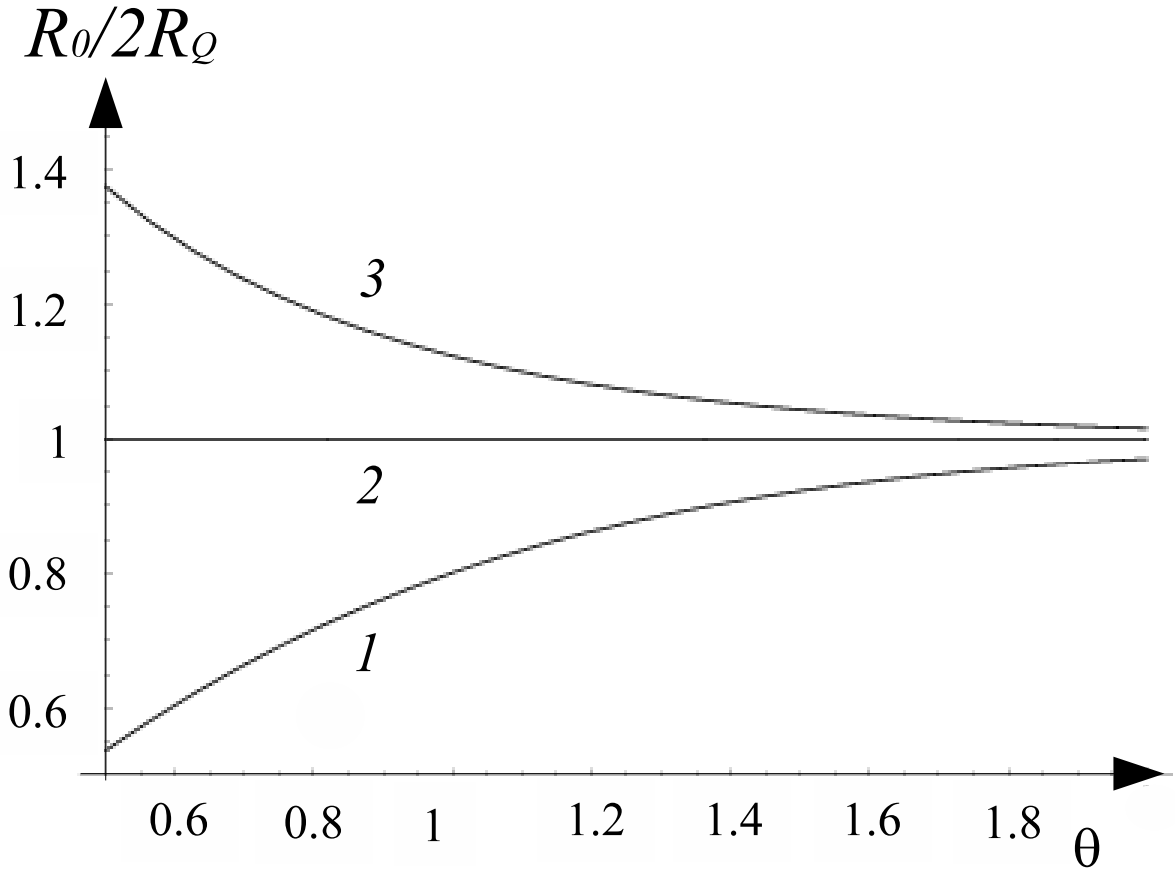}
\end{center}
\large{\caption{The resistance of the system at currents less than
the effective critical current.as a function of  $\theta$ for $b$
= 0.2 (curve 1), $b$ = 2 (curve 2), and $b$ = 5 (curve 3)   }}

\end{figure}

With account for a finite jump of the electric field at the $SN$
interface (the second term on the right in Eq. (\ref{BC2})), we
obtain for $R_{0}$

\begin{equation}
R_{0}=\frac{2}{1+r}R_{Q}\frac{2\tanh \theta +b}{2+b\tanh \theta }
\end{equation}

Consider now the case of large currents $I$ when the phase difference $%
\varphi (t)$ is increasing in time ($\varphi (t)\sim t$) and its oscillating
part is small.

\textit{a) Quasistationary case; }$I\gg I_{c}^{\ast }.$

In this case the condensate momentum is almost time-independent so that $%
\partial Q/\partial t\cong 0.$ The potentials $\mu ,V$ and the electric
field $E$ are described by Eqs.(\ref{Mux},\ref{E+V}), but the formula for
the coefficient $B$ is changed: $B=-eR_{Q}I.$ From Eq.(\ref{V_L}) we find
the voltage difference between the $N$ reservoirs

\begin{equation}
V_{NN}\equiv -2V_{L}=I[R_{B}+4R_{Q}\frac{C-1}{S}]  \label{V_NN2}
\end{equation}
and the resistance $R_{\infty }$ at large currents

\begin{equation}
R_{\infty }\equiv \lbrack R_{B}+4R_{Q}\tanh (\theta /2)]
\label{R_inf}
\end{equation}

\bigskip In the case of a long $S$ wire ($\theta \gg 1$) we obtain: $%
R_{\infty }\equiv R_{B}+4R_{Q}$; that is, the resistance of system
is the sum of the barrier resistance and the resistance of the
regions of the $S$ wire where the electric field penetrates (close
to the $SN$ interface and to the barrier). In the case of a short
$S$ wire ($\theta \ll 1$) the resistance is: $R_{\infty }\equiv
R_{B}+2\theta R_{Q},$ that is, the contribution of the
superconducting regions to the resistance decreases.

With account for the second term in Eq.(\ref{BC2}) the resistance $R_{\infty
}$ acquires the form

\begin{equation}
R_{\infty }\equiv \lbrack R_{B}+4R_{Q}(1-\frac{r}{2(1+r)})\tanh (\theta /2)]
\end{equation}
That is, the contribution of the $S$ region near the $SN$ interface to the resistance
decreases.
\section{The I-V characteristics and Shapiro steps}

In this Section we analyze the form of the current-voltage (I-V)
characteristics and calculate the voltage $V_{Sh}$ which
determines the
position of the first Shapiro step. At a finite value of the ratio $%
R_{Q}/R_{B}$ the voltage $V_{Sh}$ differs from that given by Eq.(\ref%
{JosRelation}). We restrict ourselves with the limit of large
currents $I$ employing an expansion in the parameter $I_{c}/I$. In
the considered non-stationary case all the quantities depend on
time, and in the lowest approximation in the parameter $I_{c}/I$
this dependence is determined by terms of the type $\exp (i\Omega
_{J}t)$, where $\Omega _{J}$ is the
frequency of the Josephson oscillations. Eqs.(\ref{MuEq},\ref{EqMuSt},\ref{Mux},\ref%
{E+V}) can be easily generalized for the non-stationary case. The potential $%
\mu $ is described again by the equation

\begin{equation}
\partial ^{2}\mu /\partial x^{2}=k_{\Omega }^{2}\mu  \label{EqMuNonst}
\end{equation}
where $k_{\Omega }^{2}=(i\Omega +\gamma )(i\Omega +\Omega _{\Delta
})/v_{CG}^{2}$, $ \Omega _{\Delta }=\pi \Delta ^{2}/2T$. The
solution for this equation is

\begin{equation}
\mu (x)=A_{\Omega }\cosh (k_{\Omega }x)+B_{\Omega }\sinh (k_{\Omega }x)
\label{MuxNonst}
\end{equation}

The electric field and potential have the form

\begin{equation}
eE(x)=-\frac{\Omega _{\Delta }}{i\Omega +\Omega _{\Delta }}\partial \mu
/\partial x,\text{ \ }eV(x)=\frac{\Omega _{\Delta }}{i\Omega +\Omega
_{\Delta }}\mu +eV_{\ast },  \label{E+Vnonst}
\end{equation}

In deriving Eq.(\ref{E+Vnonst}) we assume that the current $I$ does not
depend on time (no external ac signal). From the boundary conditions (\ref%
{BC1},\ref{BC2}) we find

\begin{equation}
B_{\Omega }=-eR_{\Omega }\frac{i\Omega +\Omega _{\Delta }}{\Omega _{\Delta }}%
(I-I_{c}\sin \varphi ),\text{ \ }A_{\Omega }=-(eR_{\Omega
}I(1+r)^{-1}+B_{\Omega }C_{\Omega })/S_{\Omega }  \label{B+Anonst}
\end{equation}%
where $R_{\Omega }=(k_{\Omega }\sigma )^{-1}$, $C_{\Omega }=\cosh (k_{\Omega
}L),$ $S_{\Omega }=\sinh (k_{\Omega }L)$. The electric potential at the $N$
reservoir is

\begin{equation}
eV_{L}=\frac{\Omega _{\Delta }}{i\Omega +\Omega _{\Delta }}[B_{\Omega }(\frac{%
R_{B}}{2R_{\Omega }}+S_{\Omega })+A_{\Omega }(C_{\Omega }-1)]
\label{V_Lnonst}
\end{equation}

Here we took into account the finite value of $r$. The equation
for the phase difference $\varphi $ is obtained from the
definition of $\mu _{0}=(1/4)\partial \varphi /\partial t+eV_{0}.$
It has the form

\begin{equation}
\frac{1}{2e}(\frac{\partial \varphi }{\partial t})_{\Omega
}+[R_{B}+2R_{\Omega }\frac{i\Omega +\Omega _{\Delta }}{\Omega _{\Delta }}%
\frac{C_{\Omega }}{S_{\Omega }}]I_{c}(\sin \varphi )_{\Omega }=[R_{B}+2R_{Q}(%
\frac{C_{Q}-1}{S_{Q}}+\frac{r}{(1+r)S_{Q}}]I  \label{JosEq}
\end{equation}%
\bigskip

We rearrange the terms in Eq.(\ref{JosEq}) and rewrite it in a dimensionless
form
\begin{equation}
(\frac{\partial \varphi }{\partial \tau })_{\omega }=\omega _{J}+[j(1+\frac{2%
}{b}(\frac{C_{Q}-1}{S_{Q}}+\frac{r}{(1+r)S_{Q}})-\omega _{J}]-[1+\frac{2}{%
b_{\omega }}\frac{C_{\omega }}{S_{\omega }}\frac{i\omega +\omega _{\Delta }}{%
\omega _{\Delta }}](\sin \varphi )_{\omega }  \label{JosEq1}
\end{equation}%
where $j=I/I_{c}$, $\omega _{J}=\Omega _{J}/(2eI_{c}R_{B}),$ $\tau
=t(2eI_{c}R_{B})$ are the dimensionless current, Josephson
frequency and
time, respectively; $b_{\omega }=b_{L}\theta _{\omega },$ $b_{L}=b\theta ,$ $%
\theta _{\omega }=\theta _{0}\sqrt{(i\omega +\omega _{\Delta
})(i\omega +\omega _{\gamma })}$, $\theta
_{0}=(L/v_{CG})(2I_{c}R_{B})\cong 1.11(\Delta
/T)^{3/2}\sqrt{T/E_{Th}},$ $E_{Th}=D/L^{2},$ $\omega _{\Delta
}=\Omega _{\Delta }/(2eI_{c}R_{B})=1$, $\omega _{\gamma }=\gamma
/(2eI_{c}R_{B})$.

The solution is sought in the form $\varphi (\tau )=\varphi
_{0}(\tau )+\varphi _{1}(\tau )+...,$ where $\varphi _{n}(\tau
)\sim j^{-n}$, for $n\geq 1$. In zero order approximation we
obtain from Eq.(\ref{JosEq1})

\begin{equation}
\varphi _{0}=\omega _{J0}\tau ,\text{ }\omega _{J0}=j[1+\frac{2}{b}(\frac{%
C_{Q}-1}{S_{Q}}+\frac{r}{(1+r)S_{Q}})]  \label{ZeroAppr}
\end{equation}%
The correction $\varphi _{1}(\tau )$ is equal to

\begin{equation}
\varphi _{1}(\tau )=\omega _{J0}^{-1}[\mathop{Re}\mathcal{L}(\omega )\cos
(\omega \tau )-\mathop{\ Im} \mathcal{L}(\omega )\sin (\omega \tau )]
\label{FirstCorr}
\end{equation}
where $\mathcal{L}(\omega )=[1+(2/b_{\omega })(C_{\omega }/S_{\omega
})(i\omega +\omega _{\Delta })/\omega _{\Delta }].$

Knowing $\varphi _{1}(\tau )$, one can find a correction to the
I-V curve due to the Josephson oscillations. From
Eq.(\ref{V_Lnonst}) we find for the dimensionless voltage drop
$<v_{NN}>\equiv -<2V_{L}>/(2I_{c}R_{B})$

\begin{equation}
<v_{NN}>=j[1+\frac{4}{b}(\frac{C_{Q}-1}{S_{Q}}+\frac{r}{2(1+r)S_{Q}})]-[1+%
\frac{2}{b}\tanh (\theta /2)]<\sin \varphi >  \label{v_NN}
\end{equation}

Here\ $<\sin \varphi >=<\varphi _{1}(\tau )\cos (\omega \tau
)>=(2\omega _{J0})^{-1}\mathop{Re}\mathcal{L}(\omega )$. The first
term is the Ohm's law at large currents. The second term is a
contribution from the Josephson oscillating current. Therefore the
correction to the Ohm's law is equal to

\begin{equation}
<\delta v_{NN}>=-\frac{1}{2\omega _{J0}}[1+\frac{2}{b}\tanh (\theta /2)]%
\mathop{Re}\mathcal{L}(\omega )  \label{CorrV_L}
\end{equation}%

\begin{figure}
\begin{center}
\includegraphics[width=0.9\textwidth]{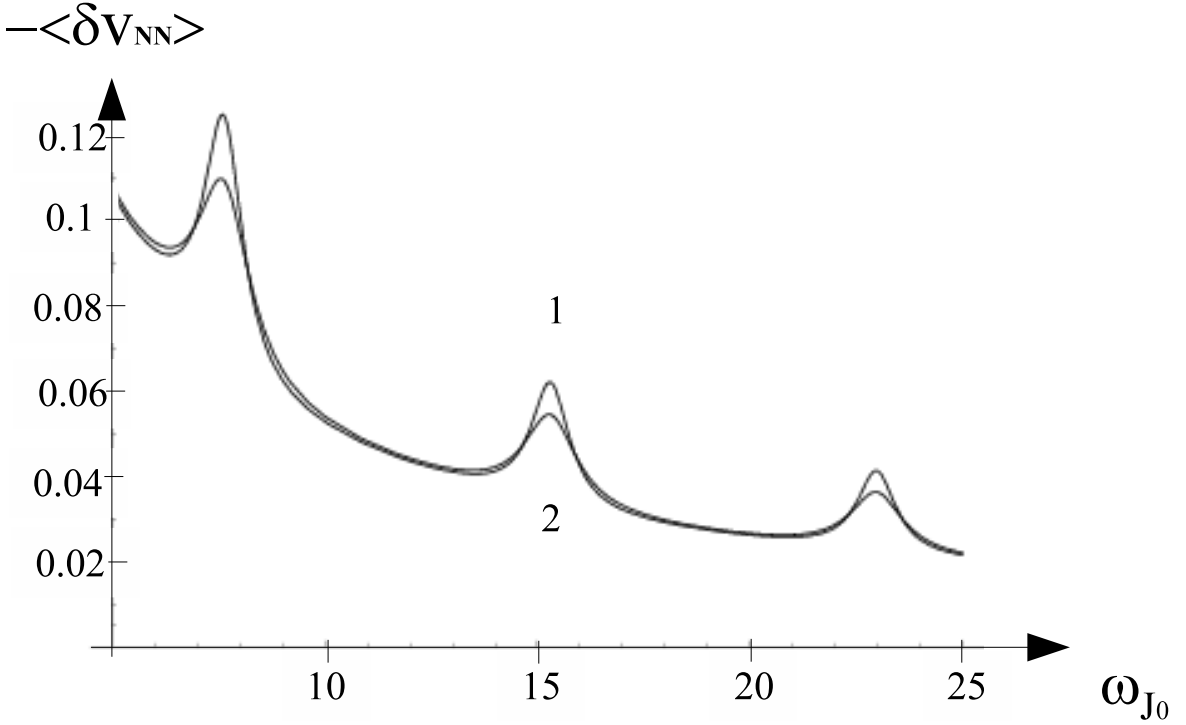}
\end{center}
\large { \caption{Correction to the I-V characteristics for
$\omega_{\gamma}$ = 0.1 (curve 1) and $\omega_{\gamma}$ = 0.5
(curve 2).}}
\end{figure}

In Fig.4 we plot the dependence of this correction $<\delta
v_{NN}>$ on the normalized frequency $\omega _{J0}$, which is
proportional to the normalized current $j$ (see
Eq.(\ref{ZeroAppr}) ) for two values of damping $\gamma $.
In plotting Fig.4 the following values of parameters are taken: $%
D=50cm^{2}/s$, $T_{c}=1.3K,$ $\Delta /T=0.3$, $b=10,$ $L=0.5mkm$,
$\tau _{\epsilon }=2\cdot 10^{-10}s.$ For these values we have:
$\omega _{\gamma }=(2eI_{c}R_{B}\tau _{\epsilon })^{-1}=(2/\pi
)(\tau _{\epsilon }T)^{-1}(T/\Delta )^{2}\cong 0.27;$ $\theta
_{0}=(\pi/2\sqrt{2})(\Delta ^{2}/T)L/\sqrt{D\Delta }\cong
1.11(\Delta /T)^{3/2}(T/E_{Th})^{1/2}\cong
0.46,$ $l_{Q}\cong 2mkm$,  $\theta =L/l_{Q}\cong 0.25,$ where $%
E_{Th}=D/L^{2}.$

It is seen that this correction has a series of peaks. These peaks
are related to excitation of a collective mode of the
Carlson-Goldman type in the system
\cite{CG,SSCollModes,AVCollModes,AVUsp,SchoenRev}. The excitation
occurs if the frequency of the Josephson oscillations $\Omega
_{J}=\omega _{J0}(2eI_{c}R_{B})$ coincides with the frequency of
the
Carlson-Goldman modes $\Omega _{CG}=v_{CG}/L$. In this case the quantity $%
\mathop{Re}\mathcal{L}(\omega )\sim \tanh ^{-1}(k_{\Omega }L)$ has a peak.
The possibility to observe such modes in another Josephson system was
studied in Ref.\cite{AVZ}.

Shapiro steps on the I-V characteristics arise if in addition to
dc current an ac current flows in the system. In the presence of
the ac current, $I_{\sim }(t)=I_{\sim }\sin (\Omega _{ex}t)$, a
new term appears on the right in Eq.(\ref{JosEq}), which is
proportional to $I_{\sim }(t)$. The position of the first Shapiro
step is determined by the equation \cite{Kulik,Barone}

\begin{equation}
\Omega _{J}\equiv (2eI_{c}R_{B})\omega _{J}=\Omega _{ex}  \label{Shap}
\end{equation}%
where $\Omega _{J}$ is the frequency of the Josephson
oscillations. We obtain the frequency of the Josephson
oscillations, $\omega _{J}$, from Eq.(\ref{JosEq1}). At large
currents $j$ in the main approximation, $\omega _{J}$ is given by
Eq.(\ref{ZeroAppr}). On the other hand, the normalized voltage
$2<v_{NN}>$\ corresponding to the same current $j$ can be easily
obtained from Eq.(\ref{V_NN2})

\begin{equation}
2<v_{NN}>=j[1+\frac{2}{b}(\tanh (\theta /2)+\frac{1}{(1+r)S})]
\label{ZeroApprV}
\end{equation}

\bigskip Therefore the deviation of $\ 2<v_{NN}>$\ from the value given by
the Josephson relation is given by the formula

\begin{equation}
\delta v_{Sh} \equiv \frac{2<v_{NN}>-\omega _{ex}}{\omega _{ex}}\mathbf{=}\frac{2}{%
(1+r)}\frac{\tanh (\theta /2)-r/S}{b+2[\tanh (\theta
/2)+r/((1+r)S)]} \label{ShapDev}
\end{equation}%
with $\omega _{ex}=\omega _{J0}$. One can see that this deviation
can be both negative and positive depending on the parameters
$\theta $\ and $r$. In the limit of large $b$\ the deviation
$\delta v_{Sh}=0$, i.e. the Josephson relation is fulfilled. At an
arbitrary $b$, the parameter $\delta v_{Sh}$\ depends on the
quantities $\theta $\ and $r$. Thus, by measuring the deviation
from the Josephson relation and the resistances $R_{0}$ and
$R_{\infty }$, one can determine the ratio $b=R_{B}/R_{Q}$, the
parameter $r$ characterizing the jump of the electric field at the
$SN$ interface, and the charge relaxation length $l_{Q}=L/\theta
.$

\section{Conclusions}

On the basis of a simple model we have analyzed the Josephson effects in a $%
NSBSN$ nanostructure at temperatures close to $T_{c}$. It turns
out that the charge imbalance taking place in this system \
essentially changes the characteristics of the Josephson junction.
If the barrier resistance $R_{B}$ is not too large in comparison
with the resistance of the $S$ wire $R_{Q}$ on the scale of the
charge imbalance relaxation length $l_{Q},$ the Josephson critical
current increases (see Eq.(\ref{EffI_c3})) and the positions of
Shapiro steps deviate from their position in equilibrium Josephson
junctions (see Eq.(\ref{ShapDev})). We have also
calculated the resistance of the system for small ($R_{0}$ for $%
I<I_{c}^{\ast }$) and large ($R_{\infty }$ for $I>>I_{c}^{\ast }$) currents.
The ratio of these resistances equals

\begin{equation}
\frac{R_{\infty }}{R_{0}}=\frac{(b+4\tanh (\theta /2))(2+b\tanh \theta )}{%
2(b+2\tanh \theta )}  \label{RatioCon}
\end{equation}

Thus, by measuring the ratio $R_{\infty }/R_{0}$ and the deviation
$\delta v_{Sh}$, one can determine the coefficient $b$ and the the
charge imbalance relaxation length $l_{Q}=L/\theta $.

We have also shown that there is a series of peaks on the $I-V$
characteristics of the system related to excitation of the
Carlson-Goldman collective mode. Note that even if the barrier
resistance $R_{B}$ is ten times larger than the resistance
$R_{Q}$, the peaks in Fig.4 caused by collective mode excitations
are clearly visible.

Note that we have adopted several assumptions. One of them is that
the voltage $V_{L\text{ }}$ should be smaller than the energy gap
$\Delta $. If the voltage is higher than $\Delta $, new effects
may arise in the system. In particular, the critical current
$I_{c}$ may change sign \cite{Klapwijk3,VolkovPRL,Zaikin,Yip}. It
would be of interest to measure the $I-V$ characteristics of the
considered system in a wide range of the applied voltages.

\section{\protect\bigskip Acknowledgements}

\bigskip I am grateful to A. Anishchanka for a technical assistance and to M.
Kharitonov for bringing Ref.\cite{Vodolazov} to my attention. I
would like to thank SFB 491 for financial support.

\section{Appendix}

Here we derive the boundary condition (\ref{BC1}) and the expression for the
current (\ref{CurrentJJ}). The latter formula in our case is not obvious
because the potential $\mu $ differs from zero and the Josephson relation (%
\ref{JosRelation}) is not valid. Consider the boundary condition at the
barrier $B$ for $4\ast 4$ matrix quasiclassical Green's function $\check{g}$ %
\cite{Zaitsev,K+L}

\begin{equation}
\sigma (\check{g}\partial \check{g}/\partial x)=(2\mathcal{S}R_{B})^{-1}[%
\check{g}_{0+,}\check{g}_{0-}]  \label{BCA}
\end{equation}
where $\check{g}_{0\pm }=\check{g}(\pm 0)$ are the Green's functions on the
right and left from the barrier $B$. The (11) and (22) elements of the $%
\check{g}$ matrix are the retarded and advanced Green's functions $\hat{g}%
^{R(A)}$ and the (12) element is the Keldysh function
$\hat{g}^{K}$. If the voltage in the system $V$ is small in
comparison with $\Delta $, the distribution function $f_{l}$ which
determines the energy gap $\Delta $ is close to the equilibrium
one ($f_{l}\approx \tanh (\epsilon /2T)$).

The current in the system is given by

\begin{equation}
I=\frac{1}{8}\mathcal{S}\sigma \int d\epsilon Tr\{\hat{\tau}_{3}[\hat{g}%
^{R}(\epsilon ,t;r)\partial \hat{g}^{K}(\epsilon ,t;r)/\partial x+\hat{g}%
^{K}(\epsilon ,t;r)\partial \hat{g}^{A}(\epsilon ,t;r)/\partial x]\}
\label{CurrentA}
\end{equation}

If we calculate the current with the help of
Eqs.(\ref{BCA},\ref{CurrentA}), we obtain Eq.(\ref{Icurrent}). Let
us calculate the current through the Josephson junction using the
right hand side of Eq.(\ref{BCA}). This current consists of three
terms: the quasiparticle current $I_{qp}$, the Josephson current
$I_{J}$ and the interference current $I_{int}$
\cite{Kulik,Barone}. The Josephson current is

\begin{equation}
I_{J}=\frac{\pi iT}{4R_{B}}\sum_{\omega }Tr\{\hat{\tau}_{3}[\hat{f}_{+}\hat{f%
}_{-}-\hat{f}_{-}\hat{f}_{+}]\}  \label{JosCurrentA}
\end{equation}
where $\hat{f}_{\pm }\equiv \hat{f}(\pm 0)=(\hat{\tau}_{2}\cos (\varphi
/2)\pm \hat{\tau}_{1}\sin (\varphi /2))\Delta /\sqrt{(\omega _{n}^{2}+\Delta
^{2})}$ are the Gor'kov's quasiclassical Green's functions on the right
(left) from the barrier. Calculating the sum, we obtain

\begin{equation}
I_{J}=I_{c}\sin \varphi  \label{I_cA}
\end{equation}
with $I_{c}=\pi \Delta ^{2}/4TeR_{B}$. The interference current is
given by

\begin{equation}
I_{int}=\frac{\cos \varphi }{16R_{B}}\int d\epsilon (\beta
V_{0})(f_{+}^{R}+f_{+}^{A}))(f_{-}^{R}+f_{-}^{A})\cosh ^{-2}(\epsilon \beta )
\label{IntCurrentA}
\end{equation}

\bigskip In the main approximation $I_{int}\approx V_{0}\Delta /(4TR_{B})%
\mathcal{S}\cos \varphi \ln (\Delta /\Gamma ),$ where $\Gamma $ is
a damping in the spectrum of the superconductor. We see that the
interference current is small in comparison with the Josephson
current if the voltage $V_{0}$ is smaller than the value $\Delta
/\ln (\Delta /\Gamma ).$ The quasiparticle current $I_{qp}$
consists of two parts: $I_{qp}=I_{qp}^{(0)}+\delta I_{qp}$, where
$I_{qp}^{(0)}$ is determined by the formula

\begin{equation}
I_{qp}^{(0)}=-\frac{1}{8eR_{B}}\int d\epsilon (\beta \partial \chi
_{+}/\partial t)(g_{+}^{R}-g_{+}^{A}))(g_{-}^{R}-g_{-}^{A})\cosh
^{-2}(\epsilon \beta )=-R_{B}^{-1}(\partial \chi _{+}/\partial t)
\label{QpCurrentA}
\end{equation}

The term $\beta \partial \chi _{+}/\partial t\cosh ^{-2}(\epsilon
\beta )$ is obtained from the equilibrium distribution function
after the
transformation of the Green's functions \cite{AVUsp}: $\check{g}\rightarrow U%
\check{g}U^{+},$ where $U=\exp (i\hat{\tau}_{3}\chi /2).$ The correction $%
\delta I_{qp}$ is determined by the response of the system to a perturbation
of the potential $\mu $ \cite{AVUsp}. It is equal to
\begin{equation}
\delta I_{qp}^{{}}=-\frac{1}{2eR_{B}}\int d\epsilon \frac{\tanh
((\epsilon
-\Omega /2)\beta )-\tanh ((\epsilon +\Omega /2)\beta )}{\Omega }\mu _{0}=%
\frac{2\mu _{0}}{eR_{B}}  \label{DeltaIqpA}
\end{equation}

where $\mu _{0}=(1/2)\partial \chi _{0}/\partial t+eV_{0}$. Thus, for the
quasiparticle current we obtain the first term in Eq.(\ref{CurrentJJ}).

The boundary condition (\ref{BCA}) may be written for the retarded
(advanced) Green's functions $\hat{g}^{R(A)}.$ Since the
distribution function $f_{l}$ approximately coincide with the
equilibrium function, one can easily obtain from this boundary
condition the equation of continuity of the condensate current at
the Josephson junction. Therefore, the quasiparticle current also
is continuous at the $SBS$ junction.


\begin{thebibliography}{99}
\bibitem{Josephson} B. D. Josephson, Rev. Mod. Phys. \textbf{46}, 251 (1974).

\bibitem{Ft1} Throughout the paper we set $\hbar =k_{B}=1$.

\bibitem{Kulik} I.O. Kulik and I.K. Yanson, Josephson effect in
superconducting tunnel structures, Keter Press, Jerusalem (1972).

\bibitem{Barone} A.Barone and G.Paterno, Physics and applications of the
Josephson effect, Wiley, NY (1982).

\bibitem{Ft2} Note that the charge imbalance is the charge of
quasiparticles. This charge is compensated by the charge of the
condensate so that the net charge is zero unless we are not
interested in small corrections of the order of $(l_{TF}/l_{Q})$,
where $l_{TF}$ and $l_{Q}$ are the Thomas-Fermi screening length
and the length of the electric field penetration into a
superconductor.

\bibitem{Cl} J. Clarke, Phys. Rev. Lett. \textbf{28}, 1363 (1972); J. Clarke
and J. L. Paterson, J. Low Temp. Phys. \textbf{15, }491 (1974);T. Y. Hsiang
and J. Clarke, Phys. Rev. B \textbf{21}, 945 (1980).

\bibitem{Yu} M. L. Yu and J. E. Mercereau, Phys. Rev. Lett. \textbf{28},
1117 (1972).

\bibitem{KlMoi} T. M. Klapwijk and J. E. Mooij, Physica \textbf{B 81}, 132
(1976).

\bibitem{Dolan} G. J. Dolan and L. D. Jackel, Phys. Rev. Lett. \textbf{39},
1628 (1979).

\bibitem{VanHar} D. J. Van\ Harlingen, J. Low Temp. Phys. \textbf{44}, 163
(1981).

\bibitem{Mamin} H. J. Mamin, J. Clarke, and D. J. Van\ Harlingen, Phys. Rev.
B \textbf{29}, 3881 (1984).

\bibitem{Chien} C.-J. Chien and V. Chandrasekhar, Phys. Rev. B \textbf{60},
3665 (1999).

\bibitem{Strunk} C. Strunk et al., Phys. Rev. B \textbf{57}, 10854 (1998).

\bibitem{Rieger} T. J. Rieger, D. J. Scalapino, and J. E. Mercereau, Phys.
Rev. Lett. \textbf{27}, 1787 (1971).

\bibitem{Tin} M. Tinkham and John Clarke, Phys. Rev. Lett. \textbf{28}, 1366
(1972); M. Tinkham, Phys. Rev.\textbf{\ B 6}, 1747 (1972).

\bibitem{SSJLTP} A. Schmid and G. Schoen, J. Low Temp. Phys. \textbf{20},
267 (1975).

\bibitem{AV} S. N. Artemenko and A. F. Volkov, JETP\ Lett. \textbf{21}, 313
(1975); Sov. Physics JETP 43, 548 (1976); S. N. Artemenko, A. F. Volkov, and
A. V. Zaitsev, J. Low Temp. Phys. \textbf{30, }487 (1978).

\bibitem{Ovch} Yu. N.\ Ovchinnikov, J. Low Temp. Phys. \textbf{31, }785
(1978).

\bibitem{AVUsp} S.N. Artemenko and A.F. Volkov, Sov. Phys. Usp. \textbf{22},
295 (1980).

\bibitem{Petrashov} V. T. Petrashov et al., Phys. Rev. Lett. \textbf{70},
347 (1993); ibid \textbf{74}, 5268 (1995).

\bibitem{Pothier} H. Pothier et al., Phys. Rev. Lett. \textbf{73}, 2488
(1994).

\bibitem{Charlat} P. Charlat et al., Phys. Rev. Lett. \textbf{77}, 4950
(1996).

\bibitem{Klapwijk} A. Dimoulas et al., Phys. Rev. Lett. \textbf{74}, 602
(1995).

\bibitem{Shapira} S. Shapira et al., Phys. Rev. Lett. \textbf{84}, 159
(2000).

\bibitem{EsteveDOS} A. Anthore, H. Pothier, and D. Esteve, Phys. Rev. Lett.
\textbf{90}, 127001 (2003).

\bibitem{SosninTherm} A. Parsons, I. A. Sosnin, and V. T. Petrashov, Phys.
Rev. \textbf{B 67}, 140502 (2003); \textit{ibid }G. Srivastava, I. Sosnin,
and V. T. Petrashov, \textbf{72}, 012514 (2005).

\bibitem{ChandraTherm} Z. Jiang and V. Chandrasekhar, Phys. Rev. Lett.
\textbf{94}, 147002 (2005); Phys. Rev. \textbf{B 72}, 020502 (2005).

\bibitem{Bezryadin} A. Rogachev, T.-C. Wei, D. Pekker, A. T. Bollinger, P.
M. Goldbart, and A. Bezryadin, Phys. Rev. Lett. \textbf{97}, 137001 (2006).

\bibitem{Feigelman} M.Yu. Kharitonov and M.V. Feigel'man, JETP Lett.,
\textbf{82}, 421 (2005).

\bibitem{Vodolazov} D. Y. Vodolazov, Phys. Rev. B \textbf{75}, 184517 (2007).

\bibitem{LO} A.I. Larkin and Yu.N. Ovchinnikov, in \textit{Nonequilibrium
Superconductivity}, edited by D.N. Langenberg and A.I. Larkin (Elsevier,
Amsterdam, 1984).

\bibitem{Rammer} J. Rammer and H. Smith, Rev. Mod. Phys. \textbf{58}, 323 (1986).

\bibitem{Kopnin} N.B. Kopnin, \textit{Theory of Nonequilibrium
Superconductivity} (Clarendon Press, Oxford, 2001).

\bibitem{Keizer}  R. S. Keizer, M. G. Flokstra, J. Aarts, and T. M.
Klapwijk, Phys. Rev. Lett. \textbf{96}, 147002 (2006).

\bibitem{Peeters}  D. Y. Vodolazov and F. M. Peeters, Phys. Rev. B \textbf{75%
}, 104515 (2007).

\bibitem{CG} P.L. Carlson and A.M. Goldman, Phys. Rev. Lett. \textbf{34}, 11
(1975).

\bibitem{SchoenRev} G. Schoen, in \textit{Nonequilibrium Superconductivity},
edited by D.N. Langenberg and A.I. Larkin (Elsevier, Amsterdam, 1984).

\bibitem{SSCollModes} A. Schmid and G. Schoen, Phys. Rev. Lett. \textbf{34},
941 (1975).

\bibitem{AVCollModes} S.N. Artemenko and A.F. Volkov, Sov. Phys. JETP
\textbf{42}, 1130 (1975).

\bibitem{Zaitsev} A.V. Zaitsev, Sov. Phys. JETP 59, 1015 (1984).

\bibitem{K+L} M.Yu. Kupriyanov and V.F. Lukichev, Sov. Phys. JETP 67, 1163
(1988).

\bibitem{AVZ} S. N. Artemenko, A. F. Volkov, and A. V. Zaitsev, JETP\ Lett.
\textbf{27}, 113 (1978).

\bibitem{Klapwijk3} J. J. A. Baselmans, A. Morpurgo, B. J. van Wees, and T.
M. Klapwijk, Nature 397, 43 (1999).

\bibitem{VolkovPRL} A. F. Volkov, Phys. Rev. Lett. \textbf{74}, 4730 (1995).

\bibitem{Zaikin} F. K. Wilhelm, G. Schoen, and A. D. Zaikin, Phys. Rev.
Lett. \textbf{81}, 1682 (1998).

\bibitem{Yip} S. K. Yip, Phys. Rev. B \textbf{58}, 5803 (1998).

\end{thebibliography}
\end{document}